\documentclass[aps,prd,twocolumn,groupedaddress,showpacs,amsmath,amssymb]{revtex4}
    \def\be{\begin{equation}}
    \def\ee{\end{equation}}
    \def\ba{\begin{eqnarray}}
    \def\ea{\end{eqnarray}}

  \input epsf
\usepackage{pstcol,graphicx,latexsym}
\usepackage{graphicx}
\usepackage{graphicx,epsf}
\usepackage{epsfig}
\begin{document}
\title{Dark Energy, Induced Gravity and Broken Scale Invariance}

\author{F. Finelli} \email{finelli@iasfbo.inaf.it}
\affiliation{INAF/IASF-BO, Istituto di Astrofisica Spaziale e 
    Fisica Cosmica di Bologna \\ Istituto Nazionale di Astrofisica \\ 
Via Gobetti 101, I-40129 Bologna -- Italy}
\affiliation{INAF/OAB, Osservatorio Astronomico di Bologna \\ Istituto 
Nazionale di Astrofisica \\
Via Ranzani 1, I-40127 Bologna -- Italy}
\affiliation{INFN, \\ via Irnerio 46, I-40126 Bologna -- Italy}
\author{A. Tronconi} \email{tronconi@bo.infn.it}
\author{G. Venturi} \email{armitage@bo.infn.it}
\affiliation{Dipartimento di Fisica, Universit\`a degli Studi di Bologna
    and INFN, \\ via Irnerio 46, I-40126 Bologna -- Italy}
\begin{abstract}
We study the cosmological evolution of an induced gravity model with a self-interacting 
scalar field $\sigma$ and in the presence of 
matter and radiation. 
Such model leads to Einstein Gravity plus a cosmological constant as a stable 
attractor among homogeneous cosmologies and is therefore a 
viable dark-energy (DE) model for a wide range of 
scalar field initial conditions and values for its positive $\gamma$
coupling to the Ricci curvature $\gamma \sigma^{2}R$. 

\end{abstract}

\maketitle
\section{Introduction}
Several years ago a model for a varying gravitational coupling 
was introduced by Brans and Dicke \cite{BD}. The model consisted of a massless scalar field whose inverse was associated with the gravitational coupling. 
Such a field evolved dynamically in the presence of matter and led to cosmological predictions differing from Einstein Gravity (EG) in that one generally 
obtained a power-law dependence on time for the gravitational coupling. 
In order to reduce such a strong time dependence in a cosmological setting, 
while retaining the Brans-Dicke results in the weak field limit, several years ago a simple model for induced gravity \cite{CV,TV} 
involving a scalar field $\sigma$ and a quartic $\lambda\sigma^{4}/4$ potential was introduced. 
This model was globally scale invariant (that is did not include any dimensional parameter) and 
the spontaneous breaking of scale invariance in such a context led to both the gravitational constant and inflation, through a non-zero cosmological constant \cite{Zee}.
The cosmological consequences of introducing matter as a perturbation were studied leading to a time dependence for the scalar field and 
consistent results. Since the present observational status is compatible with an accelerating universe dominated by dark energy (DE) 
we feel that the model should be studied in more detail. \\
In an EG framework quadratic or quartic 
potentials for canonical scalar fields are 
consistent with DE only with an extremely 
fine tuning in the initial conditions leading to slow-roll evolution until 
the present time. Indeed, a massive or a self-interacting scalar field in 
EG, respectively behave as 
dust or radiation during the coherent oscillation regime. 
In contrast with this, 
induced gravity with a self-interacting 
$\lambda\sigma^4/4$ potential has as attractor 
EG plus a cosmological constant on breaking scale invariance.
The simple model we consider illustrates how non-trivial and 
non perturbative dynamics can be obtained in the context of 
induced gravity DE, and more generally within scalar-tensor DE.

\section{The Original Model}\label{one}
Let us consider a system described by the Lagrangian
\begin{equation}
{\cal L} = {1\over 2}\sqrt{-g}\left(-g^{\mu\nu}\partial_{\mu}\sigma\partial_{\nu}\sigma+\gamma\sigma^2R-
{\lambda\over 2}\sigma^4\right)+ \sum\limits_{j=R,M} {\cal L}_{j}
\label{original}
\end{equation}
where $\gamma$, $\lambda$ are dimensionless, positive definite parameters 
and the ${\cal L}_i$'s are the contributions of cosmological fluids behaving as dust and radiation. If we consider the homogeneous mode for the scalar field evolving on a spatially flat Robertson-Walker background 
\be
ds^2 = g_{\mu \nu} dx^{\mu} dx^{\nu}=- dt^2+a^2(t) d\vec{x}^2
\ee
from the above Lagrangian one obtains the following set of independent equations
\begin{equation}\label{indepeqs}
\left\{\begin{array}{lll}
H^2=\sum\limits_{j=R,M}\frac{\rho_j}{3\gamma\sigma^2}
+\frac{1}{6\gamma}\frac{\dot \sigma^2}{\sigma^2}-2H\frac{\dot \sigma}{\sigma}
+\frac{\lambda}{12 \gamma}\sigma^2\\
\\
\frac{d}{dt}\left(a^3\sigma\dot\sigma\right)=a^3\sum\limits_{j=R,M}\frac{\left(\rho_j-3 P_j\right)}{\left(1+6\gamma\right)}\\
\\
\dot \rho_j=-3 H\left(\rho_j+P_j\right)\phantom{\frac{A}{B}}
\end{array}
\right.
\end{equation}
where $P_j=w_j\rho_j$, $w_R=1/3$, $w_M=0$, the dot denotes differentiation 
with respect to the cosmic time $t$ and $H\equiv\dot a/a$. 
Note that the system (\ref{indepeqs}) of equations are respectively 
the Friedmann, the Klein-Gordon and the continuity equations. 
In the above context, the critical densities for radiation, 
matter and $\sigma$ can be taken to be:
\be\label{cridens}
\begin{array}{ccc}
&\tilde\Omega_R = \frac{\rho_R}{3 \gamma \sigma^2 H^2} \,,\quad
\tilde\Omega_M = \frac{\rho_M}{3 \gamma \sigma^2 H^2} \,,&\\
&&\\
&\tilde\Omega_\sigma = \frac{(\dot \sigma^2 - 12 H \dot \sigma \sigma 
+\frac{ \lambda \sigma^4}{2})}{(6 \gamma \sigma^2 H^2)}\,,&
\end{array}
\ee
where $\tilde\Omega_M + \tilde\Omega_R + \tilde\Omega_\sigma =1$.\\
On using the equations of motion one finds that the curvature scalar is given by
\be
R=\frac{1}{\gamma\sigma^2}\left(\lambda \sigma^4-\dot\sigma^2+\sum\limits_{j=R,M}\frac{\rho_j-3P_j}{1+6\gamma}\right).
\label{ricci} 
\ee
Even in the absence of the fluids $\rho_j$, the quantity (\ref{ricci}) has a undefined sign, implying that the $\sigma$ potential has one or two minima depending the phase space trajectory of $\sigma$.\\
We also observe that the above model (\ref{original}) corresponds to an interacting Brans-Dicke field $\phi$. Indeed on redefining $\phi=\sigma^2$ one has
\begin{equation}
{\cal L} = - \frac{\omega}{2 \phi}
g^{\mu\nu}\partial_{\mu}\phi\partial_{\nu}\phi - \frac{\lambda}{4 \gamma^2} 
\phi^2 + \frac{\phi}{2} \, R \, + \, \sum_i L_i
\label{bdlike}
\end{equation}
where $\omega = (4 \gamma)^{-1}$. The above model (\ref{original}) also admits
a transformation in the Einstein frame in which the potential for
the redefined field is just a cosmological constant
($= 16 \pi^2 G^2 \lambda/\gamma^2$) and dark matter and
the redefined (free) scalar field are coupled (by $(\gamma^2 \sigma^4)^{-1}$).

\subsection{Super-acceleration}

Let us note that 
scalar-tensor theories admit naturally super-acceleration, i.e. 
$\dot H > 0$ (see \cite{BEPS} for a generalization of the model in \cite{CV,TV}). From the Friedmann equation and expression (\ref{ricci}) 
one finds
\begin{eqnarray}\label{suacc}
\dot H=-\frac{1}{\gamma\sigma^2}\left[\sum\limits_{j=R,M}
\frac{\left(1+8\gamma+w_j\right)}{2\left(1+6\gamma\right)}\rho_j
+ \frac{\dot \sigma^2}{2} - 4\gamma H\sigma\dot \sigma\right].
\end{eqnarray}
The possibility of super-acceleration is due to the 
$\gamma H \sigma \dot \sigma $ contribution in the brackets. 
Let us note that $\gamma$ should be large enough in order 
to have a super-accelerated phase which can be distinguished 
from a de Sitter phase. This cannot be accomplished in 
EG with a scalar field having a standard kinetic term.


\subsection{Scalar Field plus Radiation}
\begin{figure}[t!]
\centering
\epsfig{file=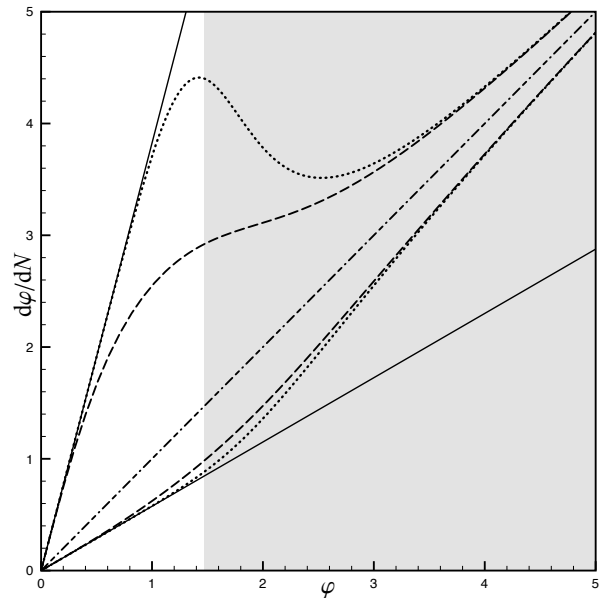, width=8 cm}
\caption{{\it Phase space plot of equation (\ref{phspeq}); on the horizontal axis  the field $\varphi$ is represented while on the vertical we plot its derivative $\frac{d\varphi}{d N}$ with $\gamma=2\cdot 10^{-1}$, $\lambda=10^{-1}$, $c_{0}=1$ and $r_{0}=0$ (dotted trajectories) or $r_{0}=5\cdot 10^{-1}$ (dashed trajectories). The solid lines are the boundaries of the accessible region of the phase space while the dotted-dashed line represents all the possible solutions with $\dot \sigma=0$. The shaded area is the portion of phase space where the scalar field has a double well potential.}\label{phasespaceplot}}
\end{figure}
When pressureless dust can be neglected with respect to radiation 
($\rho_R\gg\rho_M$) a phase space analysis can be performed for the system. 
In this case, the Klein-Gordon equation in (\ref{indepeqs}) can be 
integrated and leads to 
\be\label{kg2}
H^2=\frac{c_0}{a^8 \sigma^2 \sigma'^2}
\ee
where a prime denotes a derivative with respect to $N\equiv\ln a$ and $c_0$ is an integration constant. The Friedmann-like equation can then be cast in the form of an autonomous equation describing the possible solutions for the rescaled field $\varphi\equiv a\,\sigma$:
\begin{eqnarray}\label{phspeq}
&\gamma\, c_0^2 \left[ \varphi^2-\frac{1}{6\gamma}\left(\varphi'-\varphi\right)^2+
2 \varphi \left(\varphi'-\varphi\right) \right] =&\nonumber\\
&=\frac{\varphi^{4}}{3}
\left(\frac{\rho_{R \,, 0}}{\varphi^2}+\frac{\lambda}{4}\varphi^2\right)\,
\left(\varphi'-\varphi\right)^2&
\end{eqnarray}
where we set $\rho_{R \,, 0}=\left.\rho_R\right|_{a=1}$ (similarly we shall set
$\rho_{M \,, 0}=\left.\rho_M\right|_{a=1}$). 
The right hand side of eq. (\ref{phspeq}) is a semi-positive definite 
quantity and thus its left hand side must satisfy the inequality
\be\label{betadef}
\varphi^2\beta\equiv\varphi^2-\frac{1}{6\gamma}\left(\varphi'-\varphi\right)^2+
2 \varphi \left(\varphi'-\varphi\right)\ge 0
\ee
which determines the allowed region of phase space.\\
Once $\gamma$ and $\lambda$ are fixed, 
one must choose the initial conditions encoded 
in $c_0$ and $\rho_{R \,, 0}$ in order to determine 
completely a phase space 
trajectory. Note that if $\gamma=0$ or $c_0=0$ there is only one 
possible trajectory given by $\varphi'=\varphi$ corresponding to the GR 
limit $\dot \sigma=0$.

\section{Numerical Analysis}
If a realistic cosmological evolution is considered without any approximations, the system (\ref{indepeqs}) can only be solved numerically. On defining a rescaled Hubble parameter $\mathcal{H}=\lambda^{-1/4}H$ and introducing the dynamical variable $x=\lambda^{1/2}\sigma^2$ the relevant equations can be rewritten as
\be\label{rescsys}
\left\{
\begin{array}{lll}
\mathcal{H}^2 x''+\left[2\mathcal{H}^2+\frac{1}{2}\left(\mathcal{H}^2\right)'\right]x'=\frac{2\rho_M}{6\gamma+1}&&\\
&&\\
\mathcal{H}^2\left[1-\frac{1}{24\gamma}\left(\frac{x'}{x}\right)^2+\frac{x'}{x}\right]=\frac{\rho_M+\rho_R}{3\gamma x}+\frac{x}{12\gamma}&&
\end{array}
\right.
\ee
with $\rho_R=\rho_{R \,, 0}/a^4$ and $\rho_M=\rho_{M \,, 0}/a^3$. 
In this form the $\lambda$ dependence simplifies and one can find 
numerical solutions with a high precision even when $\lambda$ is 
extremely small which is the case (note that this is crucial since 
in the $\lambda \rightarrow 0$ limit one would otherwise just recover the 
Brans-Dicke theory).\\
The initial kinetic energy-density of the $\sigma$ field is rapidly dissipated during early stages when radiation dominates over dust, 
as can be deduced from phase space analysis and is confirmed by 
Fig. (\ref{phasespaceplot}). 
\begin{figure}[t!]
\centering
\epsfig{file=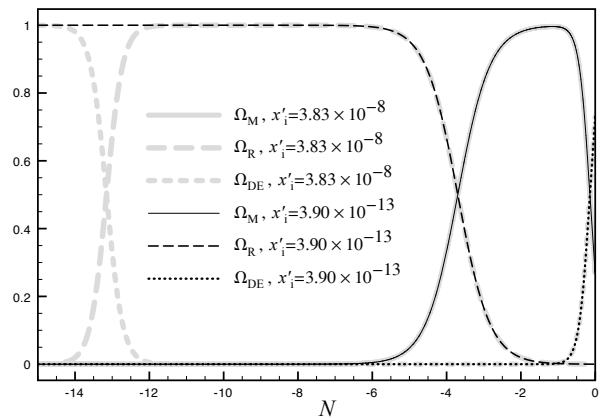, width=8 cm}
\caption{{\it Plots of 
$\Omega_R \,, \Omega_M \,, \Omega_{\rm DE}$ for the sets of 
parameters and initial conditions 1 and 2 of Table I.}
\label{densityevo}}
\end{figure}
Consideration of high values for the time derivative of the scalar field (when 
it dominates the energy-density of the universe, see Fig. \ref{densityevo}) lead to a ``stiff'' universe. 
If it existed, such an epoch should have occurred 
well before nucleosynthesis in order not to spoil the observed abundances of the 
light elements. Indeed this could be the case for very high densities arising from fermions (quarks) interacting with vector mesons (gauge fields) \cite{CV}.\\
The value for $\sigma$ remains constant until the onset of 
the matter dominated era and then its variation is well approximated by
\be\label{varsigma2}
\frac{d \sigma^2}{d N} \simeq \frac{4 \, \gamma \, \sigma_0^2}{1 + 6\gamma} \,,
\ee
where we use $\sigma_{0}=\left.\sigma\right|_{z=0}$, and then it approaches the constant value finally leading to the 
de Sitter stage. The above relation (\ref{varsigma2}) is obtained by assuming a matter (dust) dominated universe and is as expected within the context of induced gravity theories \cite{CV,BD}.

\begin{figure}[t!]
\centering
\epsfig{file=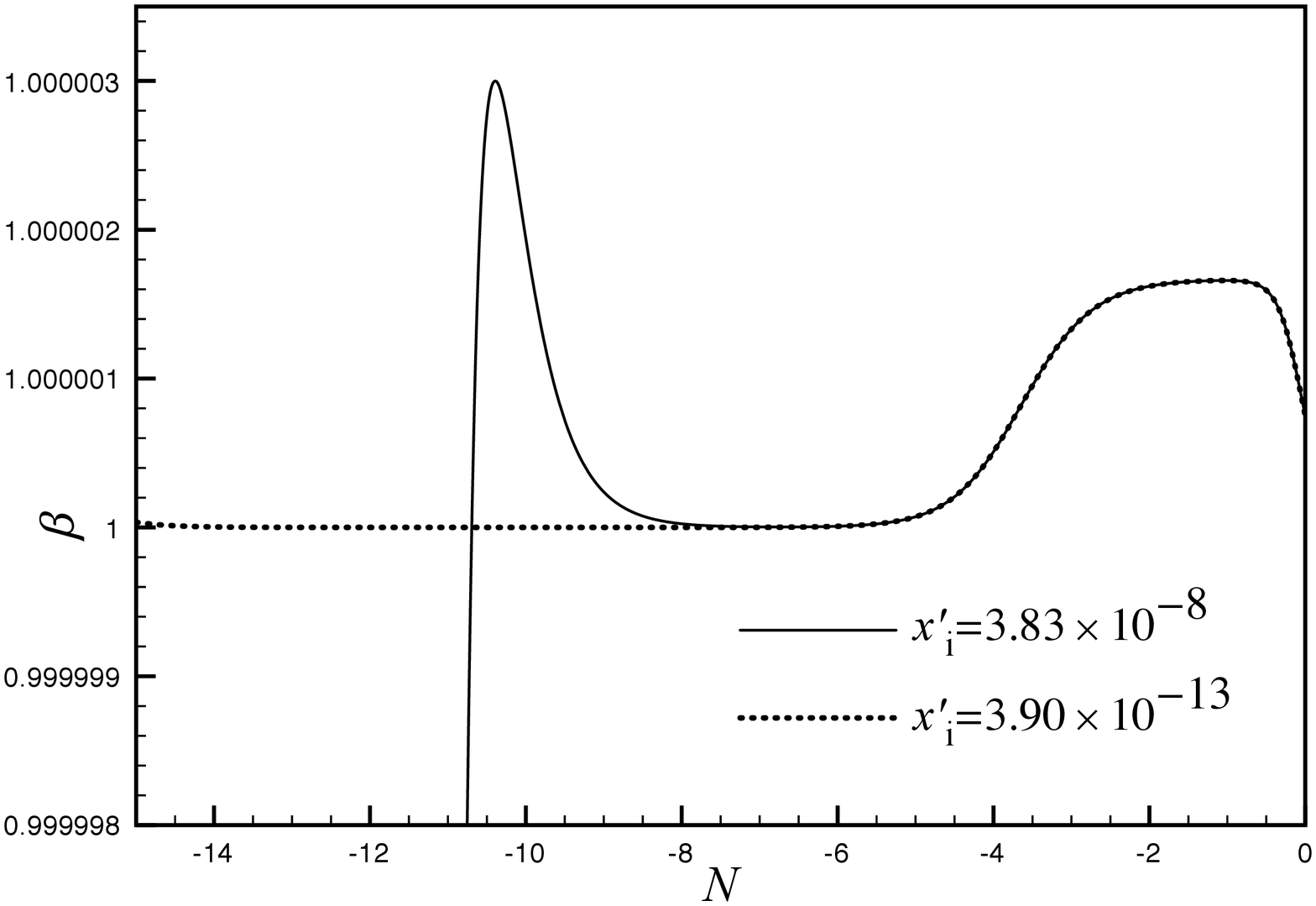, width=8 cm}
\caption{{\it Evolution of $\beta$ defined in (\ref{betadef}) plotted for the sets of initial conditions and parameters 1 and 2 of Table I.
Note that the solid and the dotted curves are indistiguishable 
between $N\sim-8$ and $N\sim 0$}
\label{betaevo}}.
\end{figure}

From here on, we shall consider
different choices of parameters and initial conditions for our analysis (see table (\ref{tab1})).
In the first two cases in table (\ref{tab1}) $\gamma$
has been taken to be its maximum value allowed by solar system constraints \cite{Bertotti:2003rm,Eubanks,gdotref}, 
while initial conditions for $\sigma$
describe two extreme situations: $\beta_{ini,1}\simeq 0$ and
$\beta_{ini,2}\simeq 1$. 
The early stages of the evolution, when the matter contribution
is negligible, are well approximated by the plots in
figure (\ref{phasespaceplot}). As expected, the field $\sigma$
rapidly evolves toward the shaded region where $\beta\simeq 1$
(see Fig. (\ref{betaevo})). Since $\gamma<<1$, when $\beta\simeq 0$ 
the Hubble parameter scales as $a^{-3}$.\\
Let us end this section by noting that one should not 
evolve the scalar field equations backwards in time - i.e. with redshift -
since the correct physical trajectory on which the scalar field 
evolves can not be identified. 

\subsection{Solar system Newtonian constraints}
The parameters of a scalar-tensor theory can be constrained 
though solar system observations which test classical general relativity 
\cite{weinberg}. 
Since our model coincides with the Brans-Dicke case in the weak field limit \cite{CV,TV}, one has \cite{weinberg,Gannouji} $\beta_{PN}-1=0$ and 
\be\label{gammaPN}
\gamma_{PN}-1=-\frac{4\gamma}{1+8\gamma}\,.
\ee
Therefore, since our $\gamma$ is positive and $\gamma_{PN}-1=(2.1\pm 2.3)\cdot 10^{-5}$ \cite{Bertotti:2003rm,Eubanks}, we shall take an upper bound for $\gamma$ given by $\gamma_{M}=5\cdot 10^{-7}$ which we shall use. Let us note however that for other values of $\gamma_{PN}$ reported in the body of Ref. \cite{Bertotti:2003rm} the resulting $\gamma_{M}$ would be 5 to 10 times larger.\\
In scalar-tensor gravity the effective Newtonian constant, namely the gravitational coupling measured 
in Cavendish like experiments, is
\be\label{Newton}
G_{{\rm eff},0}=
\frac{1}{8\pi\,\gamma\, \sigma_0^2} \frac{8\gamma+1}{6\gamma+1} \,.
\ee
Possible variations of the effective Newtonian constant $G_{\rm eff,0}$, are allowed in the interval 
$-7\cdot 10^{-14}\,y^{-1} < \dot G_{\rm eff,0}/G_{\rm eff,0} < 
3\cdot 10^{-14}\,y^{-1}$ 
\cite{gdotref}. 
The value 
$G_{N}$ of Newton's constant is known with a precision of $\sim10^{-3}$ 
\cite{Gillies:1997ff}; 
we shall thus restrict our analysis to a set of initial conditions for $\sigma$ leading to values of $G_{{\rm eff},0}$ within the bounds given by the above experimental uncertainty on the measurement of $G_{N}$.
 

\begin{widetext}

\begin{table}[t!]
\begin{center}
\begin{tabular}{|c||ccc|cc|c||}
\hline

\#&$\;\gamma/\gamma_M\;$ & $\quad x_{i}\quad$ & $\quad x'_{i}\quad$&$\quad w_{{\rm DE} \, 0}+1\quad$&$\;\chi^{2}_{snIa}\;$
&$\;(d_{{\rm L}}^{{\rm LSS}}-d_{{\rm L}\,\Lambda}^{{\rm LSS}})/d_{{\rm L}\,\Lambda}^{{\rm LSS}}\;$ \\
\hline
1& $1$ & $1.10\cdot 10^{-5}$ &$3.83\cdot 10^{-8}$&$- 10^{-6}$& $186$ & $+8.2 \cdot 10^{-6}$\\
2& $1$ & $1.12\cdot 10^{-5}$ &$3.90\cdot 10^{-13}$&$-10^{-6}$& $186$ & $-6.6 \cdot 10^{-7}$\\
3& $2^{-3}\cdot10^{2}$ & $1.056\cdot 10^{-5}$ &$1.30\cdot 10^{-7}$&$- 10^{-5}$& $186$ & $-1.1 \cdot 10^{-5}$\\
4& $2^{-3}\cdot10^{2}$ & $1.112\cdot 10^{-5}$ &$1.38\cdot 10^{-11}$&$-10^{-5}$& $186$ & $-1.4 \cdot 10^{-5}$\\
5& $2 \cdot 10^{2}$ & $1.12 \cdot 10^{-5}$ &$0$&$-2\cdot 10^{-4}$&$186$&$-1.8 \cdot 10^{-4}$\\
6& $2^{-3}\cdot10^{4}$ & $2.01 \cdot 10^{-5}$ &$-2.32 \cdot 10^{-6}$&$-10^{-3}$& $188$ & $-1.0 \cdot 10^{-3}$\\
7& $2 \cdot 10^{3}$ & $1.09 \cdot 10^{-5}$ &$0$&$-2\cdot 10^{-3}$& $189$&$-2.0\cdot 10^{-3}$\\
8& $2 \cdot 10^{4}$ & $8.32 \cdot 10^{-6}$ &$0$&$-0.02$&$234$&$-1.7 \cdot 10^{-2}$\\
\hline
\end{tabular}
\end{center}
\caption{Parameters, initial conditions and results for different 
cases. All these cases have $\Omega_M \simeq 0.27 \,, 
\Omega_{\rm DE} \simeq 0.73$, $H_0 = 73 \, {\rm km}\, {\rm s}^{-1} 
{\rm Mpc}^{-1} \,, \gamma \sigma_0^2 \simeq (8 \pi G)^{-1} \,, 
\gamma_M =5 \cdot10^{-7}$. Note that the solar system bound on $\gamma$ is relaxed for $\#$ 3 to 8; nonetheless $\#$ 3 to 7 of the model are still satisfactory on comparing with cosmological data coming from SNIa and first Doppler peak of CMBR anisotropies. The smaller the value of $\gamma$ the closer is our model to $\Lambda$CDM.} 
\label{tab1}
\end{table}

\end{widetext}

\section{Comparison with Einstein Gravity}

\begin{figure}[t!!]
\centering
\epsfig{file=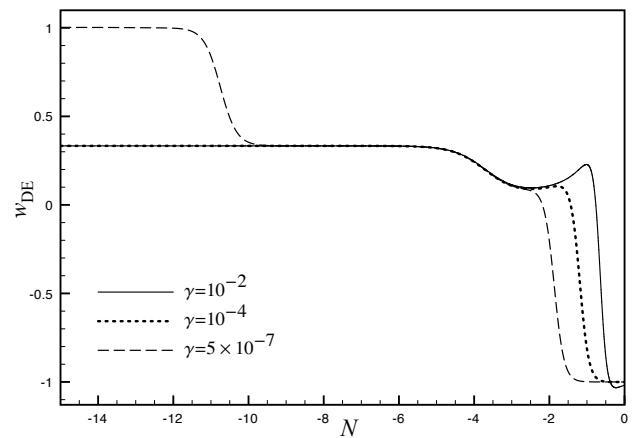, width=8.5 cm}
\caption{Evolution of $w_{\rm DE}$ as defined in (\ref{rhopfake}) for different choices of $\gamma$.\label{ws}}.
\end{figure}

In order to compare a scalar-tensor theory with 
dynamical scalar field models in EG (also dubbed quintessence 
\cite{CDS}), we first identify DE energy-density and pressure.
A consistent definition of such quantities can be given \cite{Gannouji} if one 
identifies the effective energy-density and pressure in the set of 
equations (\ref{indepeqs}) in an EG framework with 
$8\pi G_{N}\equiv\gamma\sigma_0^2$: 
\be
\left\{
\begin{array}{lll}
&&\!\!\!\!\!\!\!\rho_{\rm DE} = 3 \gamma \sigma_0^2 H^2 -\sum\limits_{j=R,M} \rho_j  \\
&&\\
&&\!\!\!\!\!\!\!\rho_{\rm DE} + p_{\rm DE} =- 2 \gamma \sigma_0^2 \dot H - \sum\limits_{j=R,M} (\rho_j + p_j) 
\end{array}
\right. \,.
\ee
Such an identification mantains the continuity equation for DE and leads 
explicitly to 
\be\label{rhopfake}
\left\{
\begin{array}{lll}
&&\!\!\!\!\!\!\!\!\rho_{\rm DE} \equiv \frac{\sigma_0^2}{\sigma^2} \left(  
\frac{\dot \sigma^2}{2}-6 \gamma H \dot \sigma \sigma + 
\frac{\lambda}{4}\sigma^4 \right) 
+ \sum\limits_{j=R,M} \rho_j \left( \frac{\sigma_0^2}{\sigma^2} 
- 1\right)\\
&&\\
&&\!\!\!\!\!\!\!\!p_{\rm DE} \equiv \frac{\sigma_0^2}{\sigma^2} \left[
\frac{\dot \sigma^2}{2}-2 \gamma H \dot \sigma \sigma 
- \frac{\lambda}{4}\sigma^4 + \sum\limits_{j=R,M} \frac{2\gamma\rho_j+p_{j}}{1+6\gamma}
\right] -p_{R}
\end{array}
\right.
\ee
We can as well identify the relative densities for the fluids involved in the evolution
\be\label{densities}
\Omega_R = \frac{\rho_R }{3 \gamma \sigma^2_0H^{2}},\,\,
\Omega_M =\frac{ \rho_M }{3 \gamma \sigma^2_0H^{2}},\,\,
\Omega_{\rm DE} =\frac{ \rho_{\rm DE}}{3 \gamma \sigma^2_0 H^{2}}
\ee
where $\Omega_M + \Omega_R + \Omega_{\rm DE} =1$.
We stress that these critical densities based on this fictitious EG model 
differ from the previous ones (\ref{cridens}) wherein
the gravitational coupling varies. For $\gamma\ll 1$ differences between the two definitions are negligible.
The matter and DE critical densities are plotted in Fig. (\ref{omegas}).  
The state parameter $w_{\rm DE}$ defined in Eq. (\ref{rhopfake})
is displayed in Fig. (\ref{ws}) for 
$\gamma =  5\times 10^{-7}\,, 10^{-4} \,, 10^{-2} $ (cases \# 1, 5, 8 in Table (\ref{tab1})). 
Note that interpreting this self-interacting 
induced gravity  as an EG model, the parameter of state of 
DE is $1/3$ during the radiation era, is positive
during the matter dominated stage (with a non negligible bump 
for $\gamma = 10^{-2}$) 
before setting to $-1+{\cal O}(\gamma)$ at present, as can be seen in 
Table I, where we have reported few examples 
of the simulations we have performed. Let us also note 
how the EG model is not completely dominated by matter, in particular 
for $\gamma \sim 10^{-2}$.\\
Since the state parameter at present 
differs only slightly from $-1$ it is therefore appropriate 
to compare this self-interacting induced gravity model with 
a $\Lambda$CDM model with $\Omega_{\rm DE \,, 0}$ 
as the fractional 
energy-density stored in the cosmological constant. 
The relative deviation of such an Einstein $\Lambda$CDM model from 
the original model for the comoving distance 
\be
d_L(z) = (1+z) \, \int_0^{z} \, \frac{d z'}{H(z')}
\label{luminositydistance}
\, ,
\ee
evaluated at the last scattering surface ($z = 1089$) 
is shown in the last column of Tab. (\ref{tab1}). The comoving distance to 
the last scattering surface enters in the acoustic scale, which is related to 
the characteristic angular scale of the peaks of the CMB angular power 
spectrum. The current uncertainty on the acoustic scale is of the order of one percent; on the basis of the numbers reported in Table  (\ref{tab1}), this implies that 
models with $\gamma \lesssim {\cal O} (10^{-2})$ may be constrained by 
the position of the peaks of the CMB angular power spectrum. 

We now compare the model with data from Supernovae Ia, 
which constrain the evolution of the scale factor for $0< z< 2$, 
as done in other works studying scalar-tensor DE \cite{NP}.
The supernova data we use are the "gold" set of 182 SNe Ia by 
Riess et al. \cite{riess}. In scalar-tensor theories, 
since the Chandrasekhar mass ($\sim x^{3/2}$) 
varies with redshift, the modulus-redshift relation is:
\be\label{magred}
\mu(z)=5\log_{10}\frac{d_{L}(z)}{10\,{\rm pc}
}+\frac{15}{4}\log_{10}\frac{x_{0}}{x(z)} \,.
\ee
In table (\ref{tab1}) the $\chi^{2}$ values are listed showing that 
$\gamma \sim 10^{-2}$ is strongly disfavoured by SNe Ia. 
An accurate statistical analysis is clearly beyond the scope of this work, 
but we expect that SNe Ia may constrain $\gamma \sim {\cal O} (10^{-3})$.

\begin{figure}[t!]
\centering
\epsfig{file=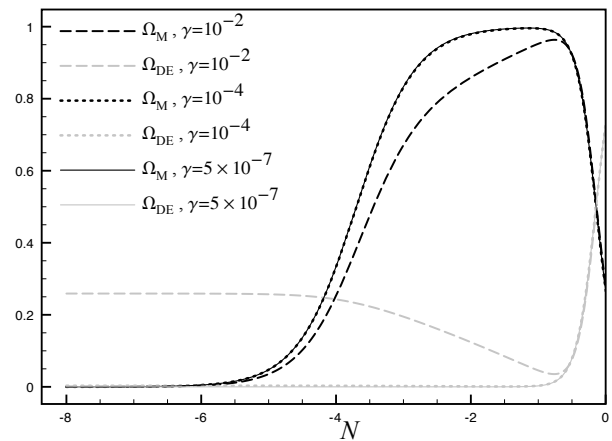, width=8 cm}
\caption{Evolution of $\Omega_{\rm M}$ and $\Omega_{\rm DE}$ as defined in (\ref{rhopfake}) for different choices of $\gamma$.\label{omegas}}.
\label{omegas}
\end{figure}
\section{Conclusions}

We have shown how a simple model of self-interacting induced gravity 
\cite{CV,TV} is a viable DE model, for tiny 
values of the self-coupling ($\lambda \sim {\cal O}(10^{-128})$). 
The model has a stable attractor 
towards EG plus $\Lambda$ and can be very similar to the 
$\Lambda$CDM model for the homogeneous mode, 
on taking into account the Solar system constraints quoted in \cite{Bertotti:2003rm,Eubanks,gdotref}.
At the cosmological level, in the presence of radiation and dust,
it is interesting that for such a simple potential (quartic for induced
gravity or interacting for the equivalent Brans-Dicke model) the model
has an attractor towards EG plus $\Lambda$, very differently from
the case of a massless scalar - i.e. $\lambda = 0$ -, for which there is
no mechanism of attraction towards EG.
The attractor mechanism towards GR and the onset to acceleration are both 
inevitably triggered by the same mechanism, i.e. scale symmetry breaking in 
this induced gravity model.
In contrast with EG, the choice of a runaway potential \cite{PR} 
for quintessence is not mandatory (see however \cite{others} for a study 
of these potential in the context of scalar-tensor theories). 
If the final attractor is an accelerated universe, the only 
constraints on parameters and initial conditions come from observations. 
We have shown that late time cosmology - after recombination, 
for instance - is mostly insensitive with respect to the initial time derivative of $\sigma$. 
For this reason, the full set of parameters and initial conditions of the 
model are fully specified on taking the observed values for 
$G, H_0, \Omega_\Lambda$.\\
We have discussed in detail such model in the context 
of Einstein gravity, i.e. keeping the Newton constant (approximately) fixed at its actual 
value. The model predicts an equation of state of the equivalent Einstein model with $w_{DE}$
slightly less than $-1$ at present and homogeneous cosmology by itself 
seems able to constrain $\gamma \lesssim {\cal O}(10^{-3})$, 
although a full statistical analysis is clearly beyond the scope of this work.\\
It is interesting to also study also what other potentials, besides the simple potential 
$\lambda \sigma^4/4$ employed in this paper, will also be compatible with the observed evolution of the universe.\\
\\
{\bf Acknowledgment.} We wish to thank the Referee for helpful and constructive criticism.




\end{document}